\documentclass[10pt]{article}
\usepackage{graphicx}
\usepackage{amsmath}
\usepackage{amssymb}
\usepackage{caption2}
\usepackage[section]{placeins}
\begin{document}

\begin{center}
{\large {\bf \sc{  Analysis of the $X(3960)$ and related tetraquark molecular states via  the QCD sum rules }}} \\[2mm]
Qi Xin${}^{*\dagger}$,
Zhi-Gang  Wang${}^*$\footnote{E-mail: zgwang@aliyun.com.},
Xiao-Song Yang$^*$

 Department of Physics, North China Electric Power University, Baoding 071003, P. R. China ${}^*$\\
 School of Nuclear Science and Engineering, North China Electric Power University, Beijing 102206, P. R. China ${}^\dagger$
\end{center}

\begin{abstract}
In this work, we study the $D\bar{D}$, $DD$, $D\bar{D}_s$, $DD_s$, $D_s\bar{D}_s$  and $D_sD_s$ tetraquark molecular states with the $J^{PC}=0^{++}$ via the QCD sum rules. The prediction $M_{D_s\bar {D}_s} = 3.98\pm0.10\, \rm{GeV}$ is in very good agreement with the experimental value  $M_{X(3960)} = 3956 \pm 5\pm 10 \,\rm{MeV}$ from the LHCb collaboration and supports assigning the $X(3960)$ as the $D_s^+D_s^-$ molecular state with the $J^{PC}=0^{++}$. We take account of our previous works on the four-quark states consisting  of two color-neutral clusters, and acquire the mass spectrum of the ground state hidden-charm and doubly-charm tetraquark molecular states.
\end{abstract}

\section{Introduction}
In 2020, the LHCb collaboration performed  an amplitude analysis of the $B^+\to D^+D^-K^+$ decays using the LHCb data taken at
the energies $\sqrt{s}=7,\,8$ and $13\,\rm{TeV}$ corresponding to a total integrated luminosity of $9\,\rm{ fb}^{-1}$, and they observed
that it is necessary to include the $\chi_{c0}(3930)$ and $\chi_{c2}(3930)$ with the $J^{PC}=0^{++}$ and $2^{++}$ respectively in the $D^+D^-$ channel, and to include the $X_0(2900)$ and $X_1(2900)$ with the $J^{P}=0^{+}$ and $1^{-}$ respectively in the $D^-K^+$ channel \cite{LHCb-X3930-PRD,LHCb-X3930-PRL}. The measured Breit-Wigner masses and widths are
\begin{eqnarray}
\chi_{c0}(3930) &:& M = 3923.8 \pm 1.5 \pm 0.4\, {\rm MeV} ;\,\,\, \Gamma = 17.4 \pm 5.1 \pm 0.8\,{\rm  MeV} \, ,\nonumber\\
\chi_{c2}(3930) &:& M = 3926.8 \pm 2.4 \pm 0.8 \, {\rm MeV} ; \,\,\, \Gamma = 34.2 \pm 6.6 \pm 1.1\, {\rm  MeV} \, .
\end{eqnarray}

Recently,  the LHCb collaboration announced the observation of the $X(3960)$ in the $D_s^+D_s^-$ invariant mass spectrum   with the significance of $12.6\,\sigma$ in the $B^{+} \to D^{+}_s D^{-}_s K^{+}$ decays, and  the assignment $J^{PC}=0^{++}$ is  favored; while the  assignments $J^{PC}=1^{--}$ and $2^{++}$ are  rejected by at least $9\,\sigma$ \cite{LHCb3960-2022}. The measured Breit-Wigner mass and width are $M = 3956 \pm 5\pm 10 $ MeV and $\Gamma = 43\pm 13\pm 8 $ MeV, respectively.

The thresholds of the $D_s^+D_s^-$ and $D^+D^-$ are $3938\,\rm{MeV}$ and $3739\,\rm{MeV}$, respectively. This leads to the possible assignments of the $X(3960)$ and $\chi_{c0}(3930)$ as the same particle.   The ratio of the branching fractions \cite{LHCb3960-2022},
\begin{eqnarray}
\frac{\Gamma(X\to D^+D^-)}{\Gamma(X\to D_s^+D_s^-)}&=&0.29\pm0.09\pm0.10\pm0.08\, ,
\end{eqnarray}
 implies the exotic nature of this state, as it is harder to excite an $s\bar{s}$ pair from the vacuum compared with the $u\bar{u}$
 or $d\bar{d}$ pair, and the conventional charmonium states predominantly decay into the $D\bar{D}$ and $D^*\bar{D}^*$ states rather than into the $D_s\bar{D}_s$ and $D_s^*\bar{D}^*_s$ states. For a short and concise review of the experimental status of the $X$, $Y$ and $Z$ states, one can consult Ref.\cite{Liu-AAPPS}.

The $D\bar{D}$ and $D_s\bar{D}_s$ molecular states have been explored via several theoretical approaches before and after the discovery of the $X(3960)$, such as the lattice QCD, the heavy quark symmetry plus light flavor $SU(3)$ symmetry, and the $SU(4)$ symmetry with breaking effects (with the couple-channel effects)  \cite{DsDs-Latt,DsDs-Heavy-SU3,DsDs-Heavy-SU3-Guo,DsDs-Heavy-Nieves,DsDs-Heavy-SU3-Hidalgo,DsDs-Heavy-SU3-Guo-after,DsDs-SU4,DsDs-SU4-Oset}.
In those theories, the $D$ and $D_s$ are physical mesons, while in the QCD sum rules, we choose the  color-singlet-color-singlet type local four-quark currents, there are two color-neutral clusters, which couple potentially to the tetraquark states with two color-neutral clusters, the color-neutral clusters have the same quantum numbers as the physical mesons except for the masses, and are compact objects, just like the diquark-antidiquark type (color antitriplet-triplet type) tetraquark states. The local currents require that they have the average spatial sizes $\sqrt{\langle r^2 \rangle}$ of the same magnitudes as the conventional mesons.

In fact, we usually choose the color antitriplet-triplet type, singlet-singlet type and octet-octet type four-quark currents to interpolate
the tetraquark states, as they couple potentially to the tetraquark states with the same quantum numbers, such as the valence quarks, $J^{PC}$, etc.
The currents with the same quantum numbers could mix with each other under re-normalization, we have to
introduce the  mixing matrixes to diagonalize those currents.
If we  choose the diagonal currents,  they  do not mix with each other and are expected to couple potentially to the physical tetraquark states, which maybe have several Fock components, such as the color antitriplet-triplet type, singlet-singlet type and octet-octet type components, etc, furthermore,  the physical masses are invariant under re-normalization. However, we cannot acquire the  mixing matrixes  without calculating the  anomalous dimension matrixes.
 It is better to choose the current having the same substructure as the main Fock component. But which component is the main Fock component? At the present time, we usually choose the criterion that if we acquire good QCD sum rules to reproduce the mass (and width) of the tetraquark state, then we reach the conclusion tentatively that its main Fock component has the same structure as the interpolating current. In the QCD sum rules, we refer the color singlet-singlet type tetraquark states as the molecular states.

In our earlier works, we have studied the mass spectroscopy of the hidden-charm and doubly-charm tetraquark (molecular) states with the QCD sum rules in a comprehensive way, and make possible assignments of the  $X$, $Y$, $Z$, $T$ states, such as the $X(3860)$, $X(3872)$, $T_{cc}(3875)$, $X(3915)$,  $X(3940)$, $X(4160)$, $Z_c(3900)$, $Z_{cs}(3985/4000)$, $Z_c(4020)$, $Z_c(4050)$, $Z_c(4055)$, $Z_c(4100)$, $Z_c(4200)$, $Z_c(4250)$, $Y(4260)$, $Y(4360)$, $Z_c(4430)$, $Z_c(4600)$,  $Y(4660)$, etc \cite{WZG-tetra-formula,WZG-Zcs3985-CPC,WZG-cc-tetra,WZG-mole-formula,WZG-hidden-mole,QXin-cc-mole}.
For more literatures on the QCD sum rules for the exotic states, one can consult the review \cite{Nielsen-review}.
In the present work, we extend our previous works on the molecular states to study the $D\bar{D}$, $D\bar {D}_s$, $D_s \bar {D}_s$, $DD$, $DD_s$, $D_s D_s$  tetraquark molecular states with the QCD sum rules and make possible assignment of the $X(3960)$ in the picture of molecular states. Furthermore, we compare our predictions to the existing works, where the $D\bar{D}$, $D\bar {D}_s$ and $D_s \bar {D}_s$ molecular states have been partly studied \cite{H.X.Chen-DDs-2020,ZhangJR-DsDs,Narison-IJMPA}. 
If the molecular states (more precisely, color-singlet-color-singlet type tetraquark states) really exist, they can decay into their constituents, through the Okubo-Zweig-Iizuka super-allowed fall-apart mechanism, saving possible in the phase-space, regardless of weak bound states or higher resonances.

The following of the article:  we obtain the QCD sum rules for the hidden-charm and doubly-charm molecular states in section 2; we offer the numerical results and discussions in section 3; conclusion is saved for section 4.

\section{QCD sum rules for  the  hidden-charm and doubly-charm molecular states}

Firstly, let us  write down  the two-point correlation functions $\Pi(p^2)$,
\begin{eqnarray}
\Pi(p^2)&=&i\int d^4x e^{ip \cdot x} \langle0|T\Big\{J(x)J^{\dagger}(0)\Big\}|0\rangle \, ,
\end{eqnarray}
where the currents $J(x)=J_{D\bar {D}}(x)$, $J_{D\bar {D}_s}(x)$, $J_{D_s \bar {D}_s}(x)$, $J_{D{D}}(x)$, $J_{D{D}_s}(x)$, $J_{D_s {D}_s}(x)$,
\begin{eqnarray}
J_{D\bar {D}}(x)&=& \bar{q}(x) i\gamma_5 c(x)\,   \bar{c}(x) i\gamma_5 q(x) \, ,\nonumber \\
J_{D\bar {D}_s}(x)&=& \bar{q}(x) i\gamma_5 c(x)\,   \bar{c}(x) i\gamma_5 s(x) \, ,\nonumber \\
J_{D_s \bar {D}_s}(x)&=& \bar{s}(x) i\gamma_5 c(x)\,   \bar{c}(x) i\gamma_5 s(x) \, ,
\end{eqnarray}
\begin{eqnarray}
J_{D{D}}(x)&=& \bar{q}(x) i\gamma_5 c(x)\,   \bar{q}(x) i\gamma_5 c(x) \, ,\nonumber \\
J_{D {D}_s}(x)&=& \bar{q}(x) i\gamma_5 c(x)\,   \bar{s}(x) i\gamma_5 c(x) \, ,\nonumber \\
J_{D_s{D}_s}(x)&=& \bar{s}(x) i\gamma_5 c(x)\,   \bar{s}(x) i\gamma_5 c(x) \, ,
\end{eqnarray}
with $q=u$, $d$. There are two color-singlet clusters in the currents $J(x)$, each has the same quantum numbers as the corresponding charmed mesons with the quantum numbers $J^{PC}=0^{-+}$. The two color-neutral clusters are in relative S-wave, thus the currents $J(x)$ have the quantum numbers $J^{PC} = 0^{++}$.

We choose the local four-quark currents, the couplings to the two-meson scattering states are neglectful \cite{WZG-Landau,WZG-comment}, we isolate the ground state tetraquark molecule contributions  to get the hadronic representation,
\begin{eqnarray}
\Pi(p^2)&=&\frac{\lambda_{X/T}^2}{M_{X/T}^2-p^2} +\cdots  \, ,
\end{eqnarray}
where the pole residues $\lambda_{X/T}$ are defined by
\begin{eqnarray}
\langle 0|J(0)|X/T(p)\rangle &=&\lambda_{X/T} \, .
\end{eqnarray}

There are two heavy quark propagators and two light quark propagators after contracting the quark fields in the correlation functions $\Pi(p^2)$ using the Wick's theorem. If each heavy quark line emits a gluon and each light quark line contributes a quark-antiquark pair, we acquire the quark-gluon operators $GG\bar{q}q \bar{q}q$, which  have dimension 10.  Therefore,  we should calculate the vacuum condensates at least up to dimension 10 to assess the convergence of the operator product expansion. We assume vacuum saturation and take account of the vacuum condensates $\langle\bar{q}q\rangle$, $\langle\frac{\alpha_{s}GG}{\pi}\rangle$, $\langle\bar{q}g_{s}\sigma Gq\rangle$, $\langle\bar{q}q\rangle^2$, $\langle\bar{q}q\rangle \langle\frac{\alpha_{s}GG}{\pi}\rangle$,  $\langle\bar{q}q\rangle  \langle\bar{q}g_{s}\sigma Gq\rangle$,
$\langle\bar{q}g_{s}\sigma Gq\rangle^2$ and $\langle\bar{q}q\rangle^2 \langle\frac{\alpha_{s}GG}{\pi}\rangle$, where $q=u$, $d$ or $s$. For the $SU(3)$ mass-breaking effects, we set $m_u=m_d=0$ and consider the terms proportional to $m_s$.

After accomplishing  the operator product expansion, we obtain the QCD representation and then reach the QCD spectral densities through dispersion relation, finally we match the hadron sides with the QCD sides, and apply Borel transform on the variable $P^2=-p^2$ to get the QCD sum rules,
\begin{eqnarray}\label{QCDSR}
\lambda^2_{X/T}\, \exp\left(-\frac{M^2_{X/T}}{T^2}\right)= \int_{4m_c^2}^{s_0} ds\, \rho_{QCD}(s) \, \exp\left(-\frac{s}{T^2}\right) \, ,
\end{eqnarray}
detailed expressions of the spectral densities $\rho_{QCD}(s)$ can be obtained by contacting the corresponding author via E-mail.

Then we obtain the masses  of the scalar molecular states through a fraction,
\begin{eqnarray}
 M^2_{X/T}&=& -\frac{\int_{4m_c^2}^{s_0} ds\frac{d}{d \tau}\rho_{QCD}(s)\exp\left(-\tau s \right)}{\int_{4m_c^2}^{s_0} ds \rho_{QCD}(s)\exp\left(-\tau s\right)}\mid_{\tau=\frac{1}{T^2}}\, .
\end{eqnarray}

\section{Numerical results and discussions}
For sake of numerical calculations, we choose the standard values of vacuum condensates $\langle\bar{q}q \rangle=-(0.24\pm 0.01\, \rm{GeV})^3$,  $\langle\bar{s}s \rangle=(0.8\pm0.1)\langle\bar{q}q \rangle$,
 $\langle\bar{q}g_s\sigma G q \rangle=m_0^2\langle \bar{q}q \rangle$, $\langle\bar{s}g_s\sigma G s \rangle=m_0^2\langle \bar{s}s \rangle$,
$m_0^2=(0.8 \pm 0.1)\,\rm{GeV}^2$, $\langle \frac{\alpha_s
GG}{\pi}\rangle=(0.33\,\rm{GeV})^4$    at the energy scale  $\mu=1\, \rm{GeV}$
\cite{SVZ-QCDSR-1979-1,SVZ-QCDSR-1979-2,Reinders-QCDSR-1985,Colangelo-QCDSR-2000}, and  take the $\overline{MS}$ masses $m_{c}(m_c)=(1.275\pm0.025)\,\rm{GeV}$
 and $m_s(\mu=2\,\rm{GeV})=(0.095\pm0.005)\,\rm{GeV}$ from the Particle Data Group \cite{PDG-2020}. Additionally, we consider the energy-scale dependence of the quark condensates, mixed quark condensates, and $\overline {MS}$ masses \cite{Narison-numberical},
 \begin{eqnarray}
 \langle\bar{q}q \rangle(\mu)&=&\langle\bar{q}q\rangle({\rm 1 GeV})\left[\frac{\alpha_{s}({\rm 1 GeV})}{\alpha_{s}(\mu)}\right]^{\frac{12}{33-2n_f}}\, , \nonumber\\
 \langle\bar{s}s \rangle(\mu)&=&\langle\bar{s}s \rangle({\rm 1 GeV})\left[\frac{\alpha_{s}({\rm 1 GeV})}{\alpha_{s}(\mu)}\right]^{\frac{12}{33-2n_f}}\, , \nonumber\\
 \langle\bar{q}g_s \sigma Gq \rangle(\mu)&=&\langle\bar{q}g_s \sigma Gq \rangle({\rm 1 GeV})\left[\frac{\alpha_{s}({\rm 1 GeV})}{\alpha_{s}(\mu)}\right]^{\frac{2}{33-2n_f}}\, ,\nonumber\\
  \langle\bar{s}g_s \sigma Gs \rangle(\mu)&=&\langle\bar{s}g_s \sigma Gs \rangle({\rm 1 GeV})\left[\frac{\alpha_{s}({\rm 1 GeV})}{\alpha_{s}(\mu)}\right]^{\frac{2}{33-2n_f}}\, ,\nonumber\\
m_c(\mu)&=&m_c(m_c)\left[\frac{\alpha_{s}(\mu)}{\alpha_{s}(m_c)}\right]^{\frac{12}{33-2n_f}} \, ,\nonumber\\
m_s(\mu)&=&m_s({\rm 2GeV} )\left[\frac{\alpha_{s}(\mu)}{\alpha_{s}({\rm 2GeV})}\right]^{\frac{12}{33-2n_f}}\, ,\nonumber\\
\alpha_s(\mu)&=&\frac{1}{b_0t}\left[1-\frac{b_1}{b_0^2}\frac{\log t}{t} +\frac{b_1^2(\log^2{t}-\log{t}-1)+b_0b_2}{b_0^4t^2}\right]\, ,
\end{eqnarray}
where $t=\log \frac{\mu^2}{\Lambda^2}$, $b_0=\frac{33-2n_f}{12\pi}$, $b_1=\frac{153-19n_f}{24\pi^2}$, $b_2=\frac{2857-\frac{5033}{9}n_f+\frac{325}{27}n_f^2}{128\pi^3}$,  $\Lambda=213\,\rm{MeV}$, $296\,\rm{MeV}$  and  $339\,\rm{MeV}$ for the quark flavors  $n_f=5$, $4$ and $3$, respectively, as the heavy quark masses play a crucial role in the QCD sum rules for the full heavy baryons, hidden(doubly)-charm(bottom) multiquark states, etc \cite{WZG-hidden-mole,QXin-cc-mole,WZG-AAPPS}.   It is preferable to select the quark flavor number $n_f = 4$ and evolve  the QCD spectral densities $\rho_{QCD}(s)$ to the desired energy scales to extract the molecule masses.  To avoid possible
failure of the energy-scale formula due to the small $D\bar{D}$($DD$) mass and achieve the most uniform parameters  we tentatively assign the $Z_c(3900)$ ($T_{cc}(3875)$) as the hidden-charm (doubly-charm)  molecular state $D \bar{D}^*+D^*\bar{D}$ ($D D^*-D^*D$) with the quantum numbers $J^{PC} = 1^{+-}$ ($J^P=1^+$), and choose the same energy scale $\mu=1.3\,\rm{GeV}$ ($1.4\,\rm{GeV}$) as in our previous work \cite{WZG-hidden-mole} (\cite{QXin-cc-mole}).

For conventional heavy mesons and quarkonia, the energy gaps between the ground states and the first radial excited states are about  $0.5\sim 0.6\,\rm{GeV}$ \cite{PDG-2020}. To avoid contaminations from the higher resonances and continuum states, we choose the continuum threshold parameters  in this work as $\sqrt{s_0}=M_{X/T}+0.4\sim 0.7\,\rm{GeV}$ and gradually adjust the continuum threshold parameters and Borel parameters via trial and error.

\begin{figure}
\centering
\includegraphics[totalheight=6cm,width=7cm]{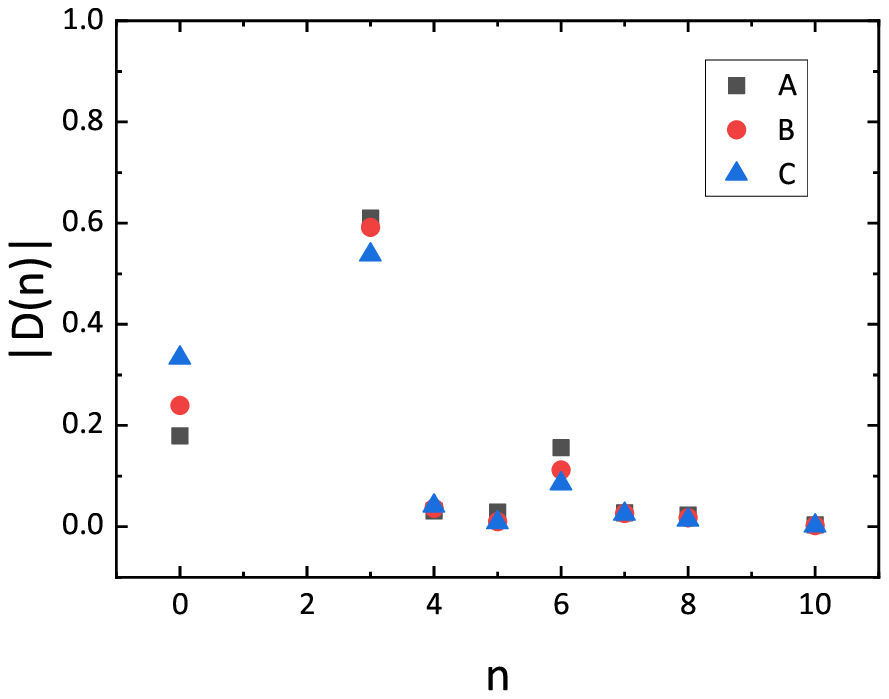}
\includegraphics[totalheight=6cm,width=7cm]{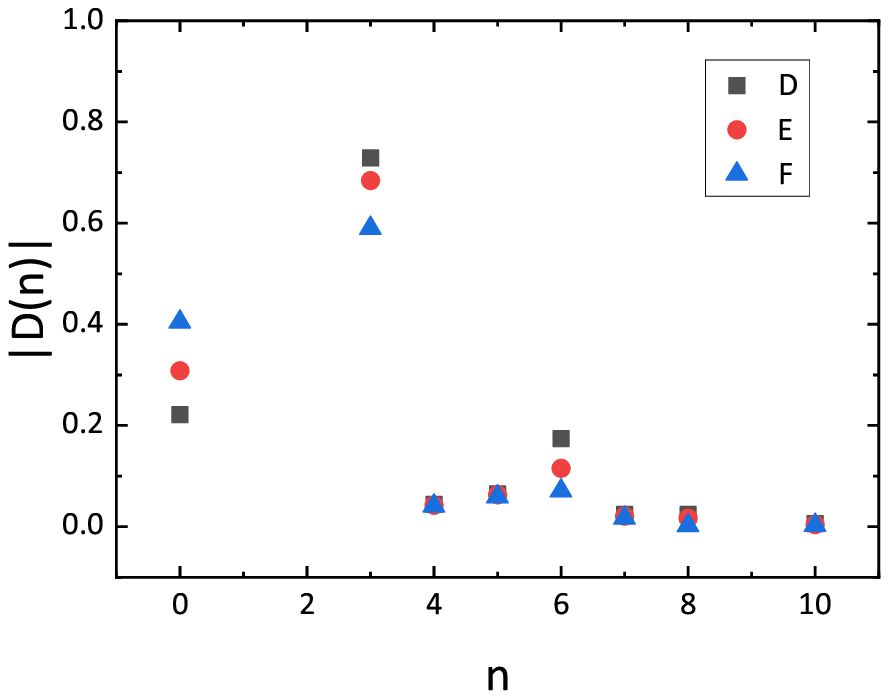}
  \caption{ The absolute values of the contributions of the vacuum condensates, where the $A$, $B$, $C$, $D$, $E$ and $F$ denote the $D\bar{D}$, $D\bar{D}_s$, $D_s \bar{D}_s$, $D{D}$, $D{D}_s$, and $D_s {D}_s$  molecular states, respectively.  }\label{dn}
\end{figure}

We obtain the Borel windows,  continuum threshold parameters and pole contributions, which are shown clearly in Table \ref{Borel-mass}. The table clearly indicates that the pole contributions  are approximately $(40-60)\%$, meanwhile the central valued are larger than $50\%$, and the expression for the pole contributions is
\begin{eqnarray}
{\rm{pole}}&=&\frac{\int_{4m_{c}^{2}}^{s_{0}}ds\rho_{QCD}\left(s\right)\exp\left(-\frac{s}{T^{2}}\right)} {\int_{4m_{c}^{2}}^{\infty}ds\rho_{QCD}\left(s\right)\exp\left(-\frac{s}{T^{2}}\right)}\, .
\end{eqnarray}
The definition of the relative contributions $D(n)$ is
\begin{eqnarray}
D(n)&=&\frac{\int_{4m_{c}^{2}}^{s_{0}}ds\rho_{QCD,n}(s)\exp\left(-\frac{s}{T^{2}}\right)}
{\int_{4m_{c}^{2}}^{s_{0}}ds\rho_{QCD}\left(s\right)\exp\left(-\frac{s}{T^{2}}\right)}\, ,
\end{eqnarray}
where the $\rho_{QCD,n}(s)$ are the spectral densities involving the vacuum condensates of dimension $n$.
The contributions $|D(10)| \ll 1\%$, and satisfy  the criterion of convergence of the operator product expansion.
In Fig.\ref{dn}, we plot the absolute values of the contributions of the vacuum condensates in the case of the central values of the input parameters. The contributions of the $|D(3)|$, both with $s$-quark and without $s$-quark,  dominate the operator product expansion,   the remaining contributions  $|D(4)|$, $|D(5)|$, $|D(7)|$, $|D(8)|$ are all $\leq4\%$, and $|D(10)|$ $\ll1\%$.

In Table \ref{Borel-mass}, we display the predicted masses and pole residues of the scalar tetraquark molecular states without strange, with strange and with hidden-strange. The prediction $M_{D_s\bar{D}_s }=3.98\pm0.10\,\rm{GeV}$ is in very good agreement with the experimental value $M = 3956 \pm 5\pm 10 $ from the LHCb collaboration and supports assigning the $X(3960)$ to be the $D_s^+D_s^-$ molecular state, the observations of the $D\bar{D}$, $DD$, $D\bar{D}_s$, $DD_s$ and $D_sD_s$  molecular states would shed light on the nature of the $X(3960)$.
The predicted mass $M_{D\bar{D}}=3.74\pm0.09\,\rm{GeV}$ lies nearby the $\psi(3770)$, which decays dominantly to the $D\bar{D}$ pair and serves as a $D$ factory,   however, the $\psi(3770)$ has the quantum numbers $J^{PC}=1^{--}$ rather than $0^{++}$, there is no overlap between the $\psi(3770)$ and $X(3740)$. On the other hand, the $\chi_{c0}(\rm 1P)$ has the mass $3414.71\pm 0.30\,\rm{MeV}$ and also has no overlap with the $X(3740)$, the observation of the $X(3740)$ plays an important role in establishing the lowest tetraquark states.  The predicted mass
$M_{D\bar{D}_s}=3.88\pm0.10\,\rm{GeV}$ is much larger than the value  $3.74 \pm 0.13\,\rm{GeV}$ from Ref.\cite{H.X.Chen-DDs-2020} due to the different schemes  in treating the QCD sum rules, such as choosing the heavy quark masses, pole contributions, etc.

In Ref.\cite{ZhangJR-DsDs}, Zhang and Huang  carry out the operator product expansion only up  to the vacuum condensates of dimension 6, and neglect the contributions of the higher dimensional condensates. They  determine the Borel windows in their scheme of the input parameters, such as the heavy quark masses and vacuum condensates, and acquire the predicted mass $M_{D_s\bar{D}_s}=3.91 \pm 0.10\, \rm{GeV}$. As the higher dimensional vacuum condensates play a great important role in obtaining the flat Borel platforms, taking into account those contributions would lead to better QCD sum rules. In Ref.\cite{Narison-IJMPA}, Albuquerque et al take account of the vacuum condensates up to dimension 8 and partly radiative $\mathcal{O}(\alpha_s)$ corrections for the perturbative terms, then choose the energy scale of the QCD spectral density  $\mu=4.5\,\rm{GeV}$  and the continuum threshold parameter $\sqrt{s_0}=6.7\,{\rm{GeV}}+2m_s(\mu)$ to extract the mass $M_{D_s\bar{D}_s}=4.169 \pm 0.048\, \rm{GeV}$. Such large energy scale $\mu$ and continuum threshold parameter $s_0$ should weaken the predictive ability.

In Tables \ref{Assignments-Table}-\ref{mass-Table-cc}, we also present  the
predictions of the molecular states involving the color-neutral clusters having the same quantum numbers as  the  $D^*$ and $D^*_s$ mesons for completeness \cite{WZG-hidden-mole,QXin-cc-mole}, all those predictions should be compared to the experimental data in the future to examine the nature of the exotic states. In Tables \ref{Borel-mass}-\ref{mass-Table-cc}, we take account of all uncertainties of the input parameters, and acquire the uncertainties $\pm 0.10\,\rm{GeV}$ and $\pm 0.09\,\rm{GeV}$ by rounding a number up or down based on the existing decimals. In fact, the difference between the uncertainties $|\delta|$ is less than $0.01\,\rm{GeV}$.

\begin{figure}
\centering
\includegraphics[totalheight=6cm,width=7cm]{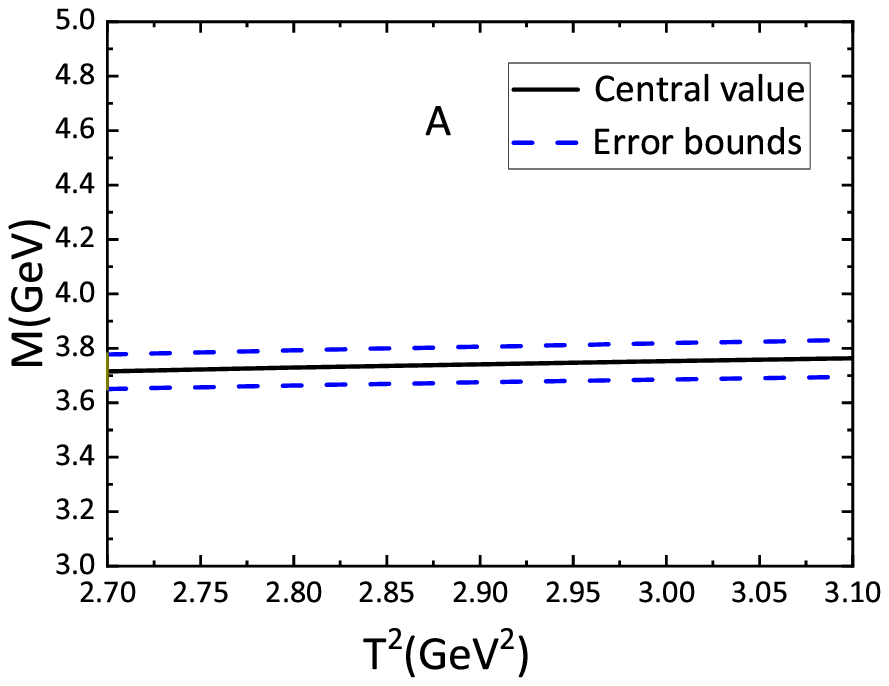}
\includegraphics[totalheight=6cm,width=7cm]{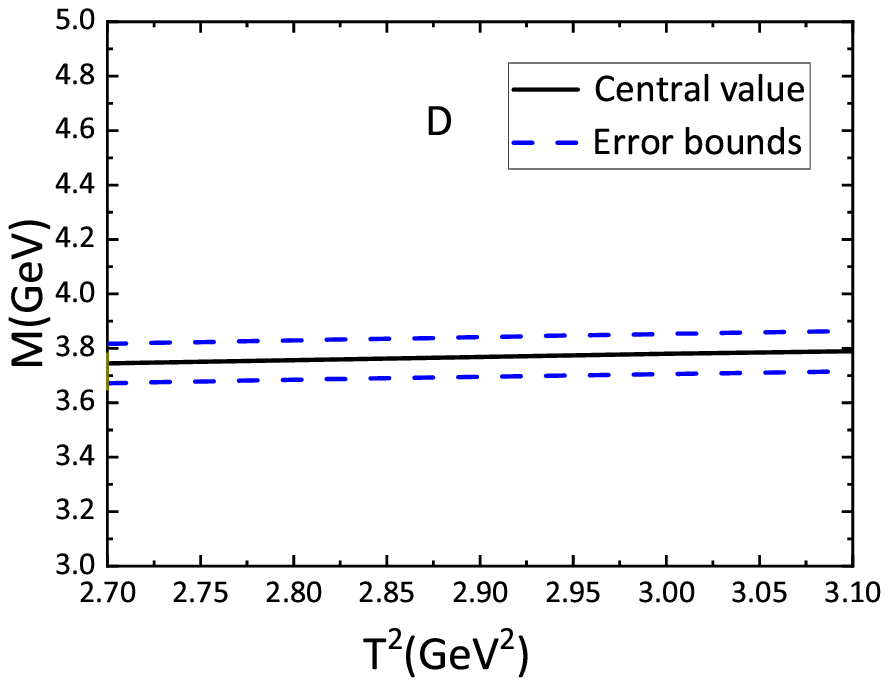}
\includegraphics[totalheight=6cm,width=7cm]{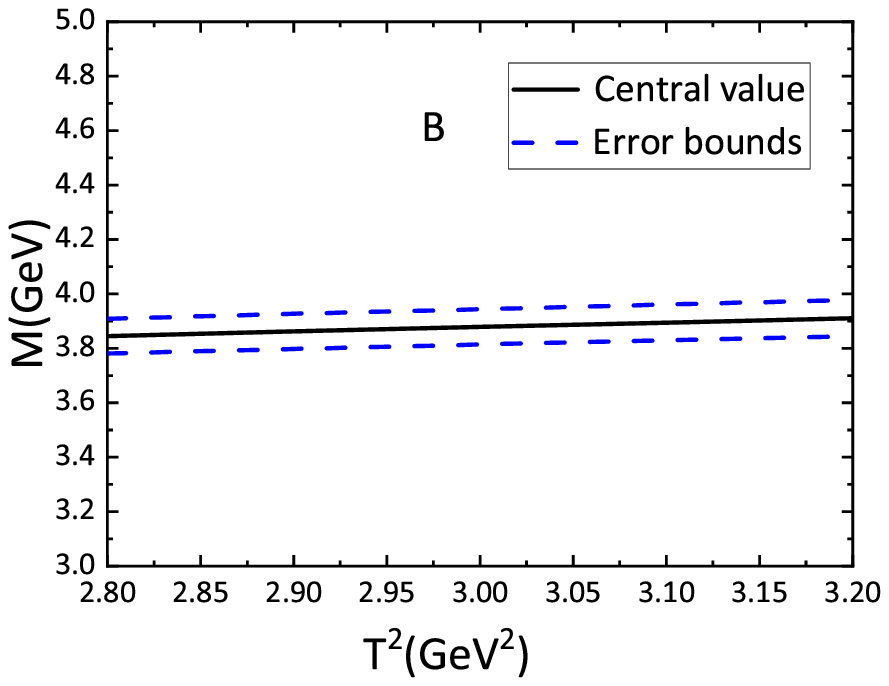}
\includegraphics[totalheight=6cm,width=7cm]{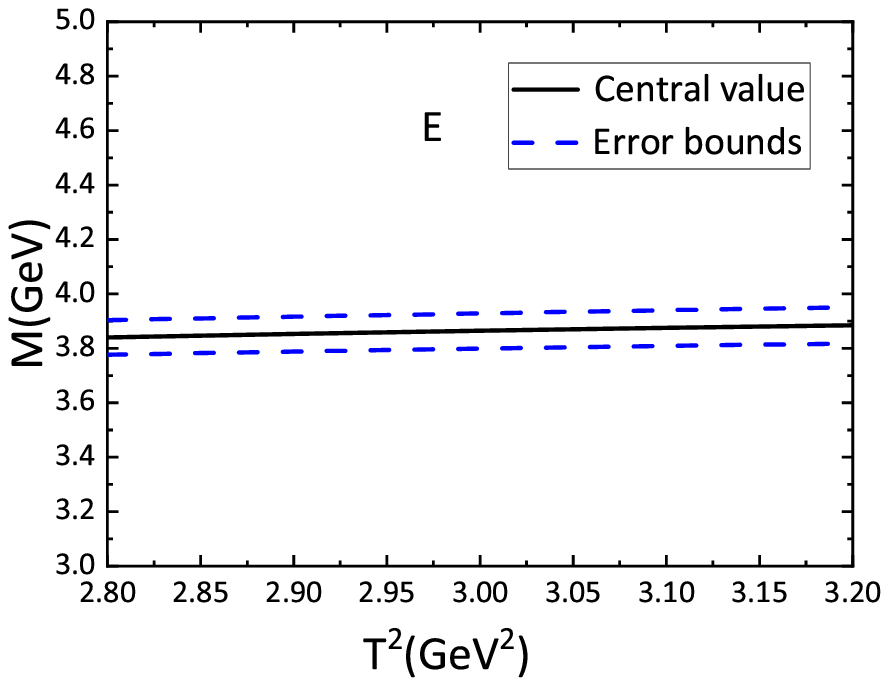}
\includegraphics[totalheight=6cm,width=7cm]{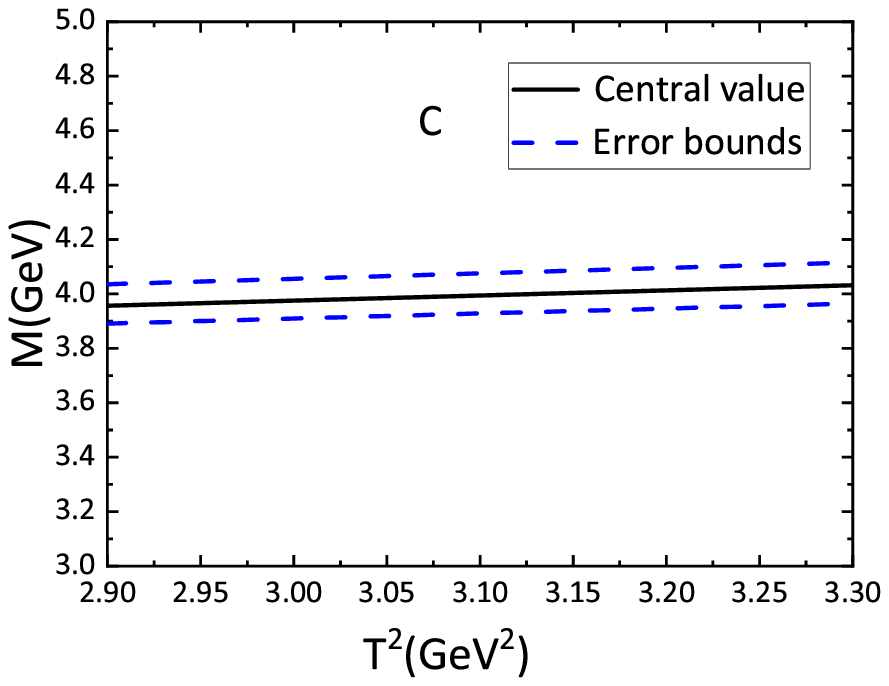}
\includegraphics[totalheight=6cm,width=7cm]{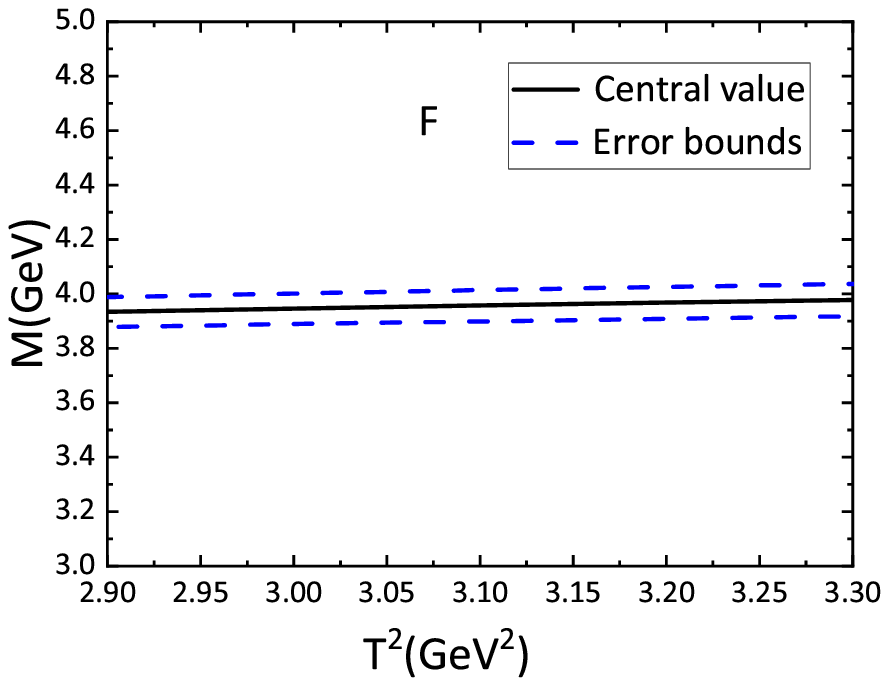}
  \caption{ The masses with variations of the Borel parameters, where the $A$, $B$, $C$, $D$, $E$ and $F$ denote the $D\bar{D}$, $D\bar{D}_s$, $D_s \bar{D}_s$, $D{D}$, $D{D}_s$, and $D_s {D}_s$  molecular states, respectively.  }\label{mass-DsP}
\end{figure}

In Fig.\ref{mass-DsP}, we plot the masses of the $D\bar {D}$, $D\bar {D}_s$,  $D_s \bar {D}_s$, $DD$, $DD_s$ and  $D_s D_s$ molecular  states according to the variations of the Borel parameters. Regardless of whether it contains $s$ quark or not, we have standardized the widths  of the Borel windows to $0.4\,\rm{ GeV}^2$ by intercepting the pole contributions $(40-60)\%$. Flat platforms emerge in the Borel windows, and the uncertainties stemming from the Borel parameters can be safely ignored, the predictions  are reliable for those molecule masses.

\begin{table}
\begin{center}
\begin{tabular}{|c|c|c|c|c|c|c|c|c|}\hline\hline
$X/T$           &$J^{PC}/J^P$&$T^2(\rm{GeV}^2)$ &$\sqrt{s_0}(\rm GeV) $ &pole          &$M_{X/T}(\rm{GeV})$ &$\lambda_{X/T}( 10^{-2}\rm{GeV}^5)$    \\ \hline

$D\bar {D}$     &$0^{++}$    &$2.7-3.1$         &$4.30\pm0.10$          &$(40-63)\%$   &$3.74\pm0.09$       &$1.61\pm0.23$     \\

$D\bar {D}_s$   &$0^{++}$    &$2.8-3.2$         &$4.40\pm0.10$          &$(40-63)\%$   &$3.88\pm0.10$       &$1.98\pm0.30$    \\

$D_s \bar {D}_s$&$0^{++}$    &$2.9-3.3$         &$4.50\pm0.10$          &$(41-62)\%$   &$3.98\pm0.10$       &$2.36\pm0.45$    \\ \hline

$D{D}$          &$0^{+}$     &$2.6-3.0$         &$4.30\pm0.10$          &$(40-64)\%$   &$3.75\pm0.09$       &$1.48 \pm0.23$     \\

$D{D}_s$        &$0^{+}$     &$2.7-3.1$         &$4.40\pm0.10$          &$(41-64)\%$   &$3.85\pm0.09$       &$1.69\pm0.26$      \\

$D_s {D}_s$     &$0^{+}$     &$2.8-3.2$         &$4.50\pm0.10$          &$(41-63)\%$   &$3.95\pm0.09$       &$2.00\pm0.32$   \\
\hline\hline
\end{tabular}
\end{center}
\caption{ The Borel parameters, continuum threshold parameters,  pole contributions,  masses and pole residues of the  molecular states. } \label{Borel-mass}
\end{table}

\begin{table}
\begin{center}
\begin{tabular}{|c|c|c|c|c|c|c|c|c|}\hline\hline
$Z_c$($X_c$)                      & $J^{PC}$   &$M_{X/Z}(\rm{GeV})$& Assignments    \\    \hline

$D\bar {D}$                       & $0^{++}$   &$3.74\pm0.09$      &               \\

$D\bar {D}_s$                     & $0^{++}$   &$3.88\pm0.10$      &              \\

$D_s \bar {D}_s$                  & $0^{++}$   &$3.98\pm0.10$      & ? $X(3960)$ \\ \hline

$D^*\bar{D}^*$                    & $0^{++}$   &$4.02\pm0.09$      &             \\

$D^*\bar{D}_s^*$                  & $0^{++}$   &$4.10\pm0.09$      &              \\

$D_s^*\bar{D}_s^*$                & $0^{++}$   &$4.20\pm0.09$      &             \\ \hline

$D\bar{D}^*-D^*\bar{D}$           & $1^{++}$   &$3.89\pm0.09$       & ? $X_c(3872)$           \\

$D\bar{D}_s^*-D^*\bar{D}_s$       & $1^{++}$   &$3.99\pm0.09$       &              \\

$D_s\bar{D}_s^*-D_s^*\bar{D}_s$   & $1^{++}$   &$4.07\pm0.09$       &             \\ \hline

$D\bar{D}^*+D^*\bar{D}$           & $1^{+-}$   &$3.89\pm0.09$       & ? $Z_c(3900)$             \\

$D\bar{D}_s^*+D^*\bar{D}_s$       & $1^{+-}$   &$3.99\pm0.09$       & ? $Z_{cs}(3985/4000)$             \\

$D_s\bar{D}_s^*+D_s^*\bar{D}_s$   & $1^{+-}$   &$4.07\pm0.09$       &               \\  \hline

$D^*\bar{D}^*$                    & $1^{+-}$   &$4.02\pm0.09$       & ? $Z_c(4020)$             \\

$D^*\bar{D}_s^*$                  & $1^{+-}$   &$4.11\pm0.09$       &      \\

$D_s^*\bar{D}_s^*$                & $1^{+-}$   &$4.19\pm0.09$       &             \\ \hline

$D^*\bar{D}^*$                    & $2^{++}$   &$4.02\pm0.09$       &                \\

$D^*\bar{D}_s^*$                  & $2^{++}$   &$4.11\pm0.09$       &              \\

$D_s^*\bar{D}_s^*$                & $2^{++}$   &$4.19\pm0.09$       &               \\
\hline\hline
\end{tabular}
\end{center}
\caption{ The possible assignments of the ground state hidden-charm  molecular states, the isospin limit is implied. The molecular states with the constituents $D^*$ and $D^*_s$ are taken from Ref.\cite{WZG-hidden-mole}. }\label{Assignments-Table}
\end{table}

\begin{table}
\begin{center}
\begin{tabular}{|c|c|c|c|c|c|c|c|c|}\hline\hline
$T_{cc}$                         &$J^{P}$           & $M_T (\rm{GeV})$   & Assignments       \\ \hline

$DD$                             &$0^{+}$           & $3.75\pm0.09$      &    \\

$D_{s}D$                         &$0^{+}$           & $3.85\pm0.09$      &   \\

$D_{s}D_{s}$                     &$0^{+}$           & $3.95\pm0.09$      &    \\           \hline

$D^{*}D^{*}$                     &$0^{+}$           & $4.04\pm0.11$      &    \\

$D_{s}^*D^{*}$                   &$0^{+}$           & $4.12\pm0.10$      &    \\

$D_{s}^*D_{s}^*$                 &$0^{+}$           & $4.22\pm0.10$      &     \\            \hline

$D^{*}D -DD^{*}$                 &$1^{+}$           & $3.88\pm0.11$      & ? $T_{cc}(3875)$ \\

$D_s^*D -D_sD^*$                 &$1^{+}$           & $3.97\pm0.10$      &   \\
\hline

$D^{*}D +DD^{*}$                 &$1^{+}$           & $3.90\pm0.11$      &    \\

$D_s^*D +D_sD^*$                 &$1^{+}$           & $3.98\pm0.11$      &    \\

$D_s^*D_s$                       &$1^{+}$           & $4.10\pm0.12$      &    \\            \hline

$D^{*}D^{*}-D^{*}D^{*}$          &$1^{+}$           & $4.00\pm0.11$      &   \\

$D_{s}^*D^{*}-D_{s}^{*}D^*$      &$1^{+}$           & $4.08\pm0.10$      &   \\

$D_{s}^*D_{s}^*$                 &$1^{+}$           & $4.19\pm0.09$      &   \\
\hline

$D^{*}D^{*}+D^{*}D^{*}$          &$2^{+}$           & $4.02\pm0.11$      &    \\

$D_{s}^*D^{*}+D_{s}^{*}D^*$      &$2^{+}$           & $4.10\pm0.11$      &    \\

$D_{s}^*D_{s}^*$                 &$2^{+}$           & $4.20\pm0.10$      &    \\
\hline\hline
\end{tabular}
\end{center}
\caption{ The possible assignments of the ground state doubly-charm  molecular states, the isospin limit is implied. The molecular states with the constituents $D^*$ and $D^*_s$ are taken from Ref.\cite{QXin-cc-mole}. }\label{mass-Table-cc}
\end{table}

\section{Conclusion}

In this work, we construct the color-singlet-color-singlet type local four-quark currents to study the $D\bar{D}$, $D\bar{D}_s$, $D_s\bar{D}_s$, $DD$, $DD_s$ and $D_sD_s$ tetraquark molecular states with the QCD sum rules. We carry out  the operator product expansion up to the vacuum condensates of dimension 10, and acquire the molecule masses after detailed analysis. The prediction $M_{D_s\bar {D}_s} = 3.98\pm0.10\, \rm{GeV}$ is in very good  agreement with the experimental value  $M = 3956 \pm 5\pm 10 $ from the LHCb collaboration and supports assigning the $X(3960)$ to be the $D_s^+D_s^-$ molecular state, the predictions for the $D\bar{D}$, $DD$,  $D\bar{D}_s$, $DD_s$ and $D_sD_s$  molecular states can be confronted to the experimental data in the future to examine the assignment of the $X(3960)$. Furthermore, we take account of  our previous works on the molecular states involving the color-neutral clusters having the same quantum numbers as the  $D^*$ and $D_s^*$ mesons, and acquire the mass spectrum of the ground state hidden-charm and doubly-charm tetraquark molecular states.

\section*{Acknowledgements}
This  work is supported by National Natural Science Foundation, Grant Number 12175068.

\end{document}